\documentclass{aa}  
\usepackage{graphicx}
\usepackage{xcolor}
\usepackage{placeins}
\usepackage{txfonts}
\usepackage[colorlinks=true,linkcolor=blue,citecolor=blue,urlcolor=blue]{hyperref}
\newcommand{\jetcaf}{{\fontfamily{qcr}\selectfont JeTCAF}}
\newcommand{\tcaf}{{\fontfamily{qcr}\selectfont TCAF}}
\newcommand{\gauss}{{\fontfamily{qcr}\selectfont GAUSSIAN}}
\newcommand{\pow}{{\fontfamily{qcr}\selectfont PL}}
\newcommand{\diskbb}{{\fontfamily{qcr}\selectfont DISKBB}}
\newcommand{\tbabs}{{\fontfamily{qcr}\selectfont TBABS}}

\newcommand{\md}{{\fontfamily{qcr}\selectfont $\dot m_{\rm d}$}}
\newcommand{\mh}{{\fontfamily{qcr}\selectfont $\dot m_{\rm h}$}}

\newcommand{\epatplot}{{\fontfamily{qcr}\selectfont epatplot}}
\newcommand{\grppha}{{\fontfamily{qcr}\selectfont grppha}}

\begin{document} 
 \title{Exploring the disc-jet scenario in 3C 273 using simultaneous {\it XMM-Newton} and {\it NuSTAR} observations}
   \subtitle{}
   \author{Ashwani Pandey\inst{1,2,3}
    \and
         Santanu Mondal \inst{4}   
    \and 
    Paul J. Wiita \inst{5}
}
 \institute{Center for Theoretical Physics, Polish Academy of Sciences, Al.Lotnikov 32/46, PL-02-668 Warsaw, Poland\\
              \email{ashwanitapan@gmail.com}
           \and  
           Department of Physics and Astronomy, University of Utah, Salt Lake City, UT 84112, USA
           \and
         Key Laboratory for Particle Astrophysics, Institute of High Energy Physics, Chinese Academy of Sciences, 19B Yuquan Road, Beijing 100049, P. R. China
           \and
           Indian Institute of Astrophysics, 2nd Block Koramangala, Bangalore 560034, Karnataka, India
           \and
           Department of Physics, The College of New Jersey, 2000 Pennington Road, Ewing, NJ 08628-0718, USA 
           }
  % \date{Received September 15, 1996; accepted March 16, 1997}

% \abstract{}{}{}{}{} 
% 5 {} token are mandatory
 
  \abstract
  % context heading (optional)
   {3C 273, a well-studied active galactic nucleus (AGN),  displays characteristics of both jetted-AGN and Seyfert galaxies, making it an excellent source to study the disc-jet connection in AGN. }  
   {To investigate the disk-jet scenario in 3C 273 using broadband (0.3--78 keV) X-ray spectra from {\it XMM-Newton} and {\it NuSTAR}. }
  % aims heading (mandatory)
   {We used  simultaneous {\it  XMM-Newton} and {\it NuSTAR} observations of 3C 273 carried out between 2012 and 2024. The 0.3--78 keV X-ray spectra were first fit with a simple power-law (PL) and then with the accretion-ejection-based \jetcaf \ model. The \jetcaf \ model accounts for emission from the jet, extending up to the sonic surface. In this framework, a reflection hump above 10 keV can also arise due to the bulk motion Comptonization of coronal photons by the jet.
   }
  % methods heading (mandatory)
   {We found that the simple PL did not provide a good fit, leaving significant residuals at energies below 1.5 keV. All the spectra were fitted well by the \jetcaf \ model. The weighted-averaged black hole mass of (7.77$\pm$0.30) $\times 10^8 M_\odot$ obtained from the \jetcaf \ model is comparable with the previous estimates based on reverberation mapping observations and accretion disk models. }
  % results heading (mandatory)
   {The 0.3--78 keV X-ray emission of 3C 273 can be fit by the accretion-ejection-based model in which the corona and the jet on top of it make significant contributions to the X-ray flux. The Doppler boosting factor estimated from the jet flux ranges from 1.6 to 2.2, consistent with the lower limit from the literature.}
  % conclusions heading (optional), leave it empty if necessary 
   {}

   \keywords{Accretion, accretion disks -- Galaxies: active -- Galaxies: jets -- X-rays: individuals (3C 273)}
\titlerunning{Exploring the disc-jet scenario in 3C 273}
   \maketitle
%
%----------------------------- Introduction --------------------------------------

\section{Introduction}\label{sec:intro}
It is now well established that the nuclei of active galaxies (called active galactic nuclei or AGN) host supermassive black holes that actively accrete material through a disc and generate electromagnetic radiations \citep[e.g.][]{1984ARA&A..22..471R}. Evidence of highly collimated relativistic outflows, or jets, has been found in a fraction ($\sim10-20$\%) of AGN, known as jetted-AGN, from their high-resolution radio imaging and multi-wavelength observations \citep[e.g.][]{2019ARA&A..57..467B}. The formation, acceleration, and collimation of these relativistic jets are still not fully understood. Models available in the literature for jet production fundamentally assume that they originate in the vicinity of the black holes and extract their power mainly from (a) the black hole spin \citep{1977MNRAS.179..433B}, and/or  (b) the accretion disc \citep{1982MNRAS.199..883B}.  Both of these basic models expect a connection between the relativistic jet and the accretion disc. The disc-jet connection in accreting systems is one of the most important unresolved issues in astrophysics and has been the focus of many investigations \citep[e.g.][]{2003ApJ...593..667M,2003ApJ...593..184L,2014MNRAS.445...81S,2019MNRAS.486.1672M}. 

AGN having relativistic jets aligned closely to the observer are known as blazars \citep[e.g.][]{2017A&ARv..25....2P}. 
Based on the equivalent widths (EW) of emission lines in their optical spectra, \cite{1991ApJS...76..813S} classified them as Flat Spectrum Radio Quasars (FSRQs; EW$_{rest}$ $>5$\AA) and BL Lacertae objects (BLLs; EW$_{rest}$ $< 5$\AA). The absence of strong emission lines in BLLs could be due to the presence of radiatively inefficient disc \citep{2011MNRAS.414.2674G}. 

3C 273, the first identified quasar, is a nearby  \citep[$z = 0.158$;][]{1963Natur.197.1040S} highly luminous FSRQ. It is extremely variable across all the electromagnetic (EM) frequencies \citep[e.g.][]{2008A&A...486..411S}; however, unlike most other FSRQs it shows a low (on average $<1$ \%) degree of optical polarization \citep{1991AJ....102.1946V,2018A&A...620A..68H}. Due to its high luminosity and proximity, it has been intensively monitored for flux and spectral variability over the entire (that is, from radio to $\gamma-$rays) EM spectrum \citep[e.g.][]{1999ApJ...522..846X,2000A&A...361..850T,2001ApJ...549L.161S,2002MNRAS.336..932K,2009AJ....138.1428F,2014ApJS..213...26F,2010ApJ...714L..73A,2015MNRAS.451.1356K,2015ApJ...812...14M,2016A&A...590A..61C,2020MNRAS.497.2066F}. 

The submillimetre-to-radio emission of 3C 273 is characterised by strong flux variations that are produced by synchrotron emission of relativistic electrons within the jet \citep{2000A&A...361..850T,2008A&A...486..411S}. At optical-to-UV frequencies, a bright excess (blue-bump) is usually found that can be interpreted as the contribution from two differently variable components; a blue component and a red component \citep{1998A&A...340...47P}. The blue component, which is mostly variable in the UV, can be attributed to thermal emission from the accretion disc, while the red component, which is significantly variable in the IR, can be due to the jet emission \citep{2008A&A...486..411S}. 

In the X-ray band, a soft excess is commonly observed in the low energy (below $\sim$2 keV) spectra of 3C 273 that can be explained by the thermal Comptonization of UV disc photons in a hot corona above the disc \citep{Grandi2004}. A correlation between low-energy X-ray and UV emission has been found in a few observations that support the Comptonization scenario \citep[e.g.][]{1992A&A...258..255W,2015MNRAS.451.1356K}. However, such a correlation was not detected in certain studies \citep[e.g.][]{2007A&A...465..147C,2008A&A...486..411S} that question this interpretation.

The spectra of such `jetted' sources can be described using accretion disc-jet-based models \citep[][and references therein]{WandelUrry1991,ZdziarskiGrandi2001,Grandi2004,MondalEtal2022A&A...663A.178M,DasChatterjee2023MNRAS.524.3797D}, where UV and X-ray emission may come from the vicinity of the accretion disc. These works found the signature of an accretion disc along with the jet in the X-ray spectra of blazars and FSRQs. As 3C 273 is one such candidate, we attempt to further investigate this possibility by fitting simultaneous broadband X-ray observations from {\it NuSTAR}  and {\it XMM-Newton} using an accretion-ejection-based two-component advective flow \citep[\tcaf;][]{ChakrabartiTitarchuk1995ApJ...455..623C} model that has been expanded to include jet emission  \citep[\jetcaf;][]{MondalChakrabarti2021}. The \jetcaf\, model takes into account the radiation mechanisms in the disk, corona, at the base of the jet/outflows and the effect of bulk motion by the outflowing jet on the emitted spectra. We note that apart from the base of the jet, the rest of the jet can also contribute to the spectrum. In the present model, the jet is only considered up to the sonic surface. If the inclusion of the rest of the jet does not change the overall spectral shape but only the total flux in X-rays,
the present model parameters could fit the contribution from the rest of the jet with some changes in the model normalization. This model has six parameters, namely: (i) the mass of the BH ($M_{\rm BH}$) if it is unknown; (ii) the Keplerian disk accretion rate ($\dot m_{\rm d}$); (iii) the sub-Keplerian halo accretion rate ($\dot m_{\rm h}$); (iv) the size of the dynamic corona or the location of the shock within the accretion flow \citep[$X_{\rm s}$ in units of $r_S=2GM_{\rm BH}/c^2$;][]{Chakrabarti1989ApJ...347..365C}; 
(v) the shock compression ratio ($R$) which is the velocity drop across the shock, therefore the jump in density there; and (vi) the ratio of the solid angle subtended by the outflow to the inflow, $f_{\rm col}(\equiv \Theta_o/\Theta_{\rm in}$). The final parameter may depend on the jet properties, however, since we do not know them beforehand, we take it as a user-defined parameter. Since the mass of the BH is also a parameter in this model, one can determine its value from spectral analysis\citep[e.g.,][]{MondalEtal2022A&A...662A..77M}, as was done by using the \tcaf\,model fitting within the standard software package, XSPEC \citep{Debnathetal2014,MollaEtal2017ApJ...834...88M}. The spectrum of 3C 273 is complex, including disc and power-law components, soft excess, and reflection signature. The \jetcaf\, model incorporates both the corona to disc and disc to corona photon interceptions to iteratively compute the spectrum with a modified disc temperature similar to \tcaf. In addition, when Comptonized photons from the corona pass through the jet medium, this produces a Compton hump. The combination of these two processes changes the X-ray spectrum significantly by producing a hump in above 10 keV, similar to so-called reflection models. However, the current model does not include emission processes and therefore can not produce the often observed iron lines. 

The structure of this paper is as follows. In Section \ref{sec:obs_data} we describe the observations and data reduction procedure. The results and our discussion of them are presented in Section \ref{sec:result}. Section \ref{sec:conclusion} presents our conclusions. 
\begin{table*}
\scriptsize
\centering
\caption{\label{table:obslog}A log of simultaneous {\it NuSTAR} and {\it XMM-Newton} observations.} 
%\scalebox{1.2}{
\begin{tabular}{cccccc}
\hline%\hline
\multicolumn{3}{c}{\textbf{\textit{NuSTAR}}} & \multicolumn{3}{c}{\textbf{\textit{XMM-Newton}}}  \\\hline
ObsID        &   Observation start date and time & Exposure (ks) & ObsID        &   Observation start date and time & Exposure (ks)   \\
10012004001  &	2012-07-13 11:11:07  & 6.23   &	     -       &           -             &    -     \\
10002020001  &	2012-07-14 00:06:07  & 243.97 &	0414191001  &	2012-07-16 11:59:23  &	38.92  \\
10002020003  &	2015-07-13 14:01:08  & 49.41  &	0414191101  &	2015-07-13 21:03:55  &	72.40  \\
10202020002  &	2016-06-26 19:11:08  & 35.41  &	0414191201  &	2016-06-26 20:22:08  &	67.20  \\
10302020002  &	2017-06-26 17:41:09  & 35.40  &	0414191301  &	2017-06-26 19:15:23  &	67.00  \\	
80301602002  &	2018-05-19 18:01:09  & 60.76  &     -        &          -              &    -     \\
10402020002  &	2018-06-02 02:01:09  & 16.09  &	    -        &          -              &    -     \\
10402020004  &	2018-06-15 04:31:09  & 21.18  &	    -        &          -              &    -     \\
10402020006  &	2018-07-04 17:21:09  & 40.32  &	0414191401  &	2018-07-04 17:54:16  &	78.00  \\
10502620002  &	2019-07-02 07:51:09  & 49.41  &	0810820101  &	2019-07-02 17:16:32  &	69.40  \\
10602606002  &	2020-07-06 04:56:09  & 44.02  & 0810821501  &	2020-07-06 11:59:20  &	69.90  \\
60601004002  &	2021-04-14 14:46:09  & 18.67  &     -        &           -             &   -      \\
10702608002  &	2021-06-09 18:36:09  & 36.04  &	0810821601  &	2021-06-09 19:26:58  &	65.00  \\
10802608002  &  2022-06-28 01:11:09  & 33.18  & 0810821901  &	2022-06-28 02:36:00  &	61.80  \\
60701019002  &  2023-04-12 17:16:09  & 19.75  &    -        &           -            &    -    \\
11002608002  &  2024-01-07 11:16:09  & 43.03  & 0810822101  &   2024-01-07 16:43:21  &	62.99  \\
\hline
\end{tabular}
	%}
\end{table*}
%----------------------------- Observations and data reduction ------------------------------------------------
\begin{figure*}
    \centering
     \includegraphics[width=8cm, height=6cm]{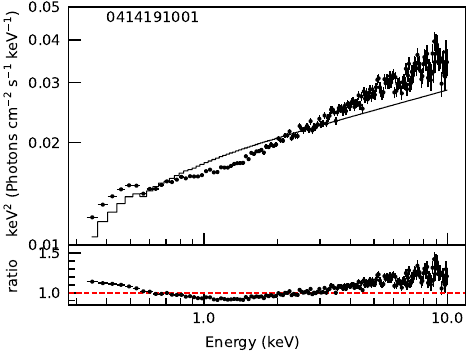} \hfill \includegraphics[width=8cm, height=6cm]{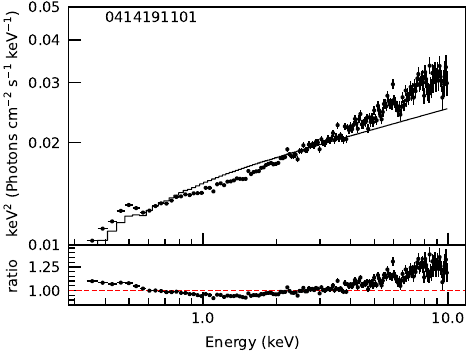}\\
    \caption{\label{fig:ratio_PL} Sample plots showing \tbabs(\pow) fitted 0.3-10 keV spectra of 3C 273 (top panel) and the ratio of data/model (bottom panel). The observation ID is mentioned in each plot. Plots for the remaining {\it XMM-Newton} observations are shown in Figure \ref{fig:pl_spec} in the appendix.}
\end{figure*}

\begin{figure*}
    \centering
    \includegraphics[width=8cm, height=6cm]{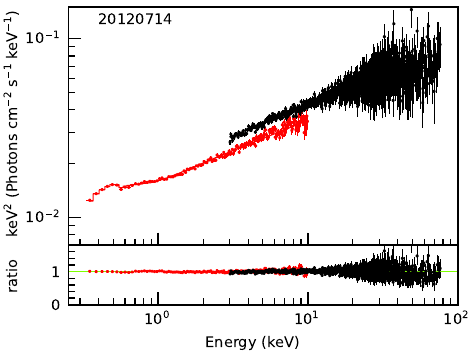}\hfill
     \includegraphics[width=8cm, height=6cm]{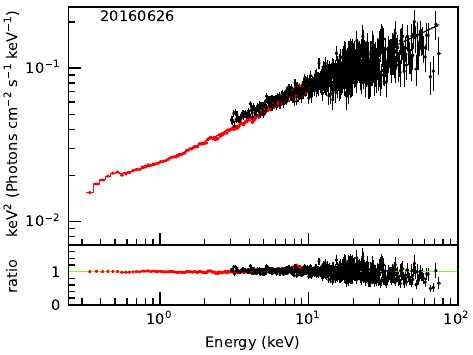} \\
    \caption{\label{fig:2} Sample plots showing \tbabs(\diskbb+\pow) fitted 0.3-78 keV spectra of 3C 273 (top panel) and the ratio of data/model (bottom panel). The observation date (yyyymmdd) is mentioned in each plot. Plots for the remaining simultaneous observations are shown in Figure \ref{fig:diskbb+pl_spec} in the appendix.}
\end{figure*}

\begin{table*}
\centering
\caption{\label{table:disk_fit} Best fitted \tbabs(\diskbb+\pow) model parameters for all simultaneous {\it XMM-Newton} and {\it NuSTAR} observations of 3C\,273. }
\begin{tabular}{cccccc}
\hline
Obs Date &    MJD       & T$_{\rm in} (keV)$ & Norm$_{diskbb}$ &   $\Gamma_{\rm PL}$ & $\chi_{\rm r}^2$ \\ \hline
2012-07-14 &   56122    & 0.131 $\pm$ 0.003 & 2996.57 $\pm$ 353.90  &  1.664 $\pm$ 0.003  & 1.10 \\ 
2015-07-13 & 57216	& 0.142 $\pm$ 0.004 & 1467.65 $\pm$ 190.93  &  1.680 $\pm$ 0.005  & 1.22 \\ 
2016-06-26 & 57565	& 0.136 $\pm$ 0.003 & 3228.20 $\pm$ 304.37  &  1.540 $\pm$ 0.004  & 1.57 \\ 
2017-06-26 & 57930	& 0.137 $\pm$ 0.003 & 1409.50 $\pm$ 164.90  &  1.587 $\pm$ 0.004  & 1.25 \\ 
2018-07-04 & 58303	& 0.137 $\pm$ 0.003 & 2646.79 $\pm$ 228.69  &  1.689 $\pm$ 0.004  & 1.37 \\ 
2019-07-02 & 58666	& 0.149 $\pm$ 0.003 & 1700.25 $\pm$ 131.37  &  1.719 $\pm$ 0.005  & 1.34 \\ 
2020-07-06 & 59036	& 0.131 $\pm$ 0.002 & 3779.85 $\pm$ 292.15  &  1.660 $\pm$ 0.004  & 1.43 \\ 
2021-06-09 & 59374	& 0.146 $\pm$ 0.002 & 1979.40 $\pm$ 151.85  &  1.665 $\pm$ 0.006  & 1.44 \\ 
2022-06-28 & 59758	& 0.134 $\pm$ 0.002 & 3382.95 $\pm$ 275.60  &  1.682 $\pm$ 0.005  & 1.55 \\ 
2024-01-07 & 60316	& 0.129 $\pm$ 0.004 & 1917.43 $\pm$ 283.80  &  1.630 $\pm$ 0.005  & 1.21 \\ 
\hline
\end{tabular} 
\end{table*}
\section{Observation and data analysis}\label{sec:obs_data}
{\it NuSTAR} observed the blazar 3C 273 on 32 occasions between 2012 July 1 and 2024 January 7. In this work, we only selected observations with exposure time greater than 5 ks. Also, on 2012 July 13, out of the six observations of almost equal exposures ($\sim$ 6 ks), we only used the one observation with the largest exposure time. 
Finally, we are left with a total of 16 observations with an exposure time ranging from 6.23 ks to 243.97 ks that were performed between 2012 July 13 and 2024 January 7. A detailed log of these observations is given in Table \ref{table:obslog}.

We downloaded the {\it NuSTAR} observations of 3C 273 from the HEASARC Data Archive\footnote{\url{https://heasarc.gsfc.nasa.gov/cgi-bin/W3Browse/w3browse.pl}}. We followed the standard procedures\footnote{\url{https://heasarc.gsfc.nasa.gov/docs/nustar/analysis/}} to reduce and analyse the {\it NuSTAR} data sets using HEASOFT version 6.29 and CALDB version 20210427. We first generated the calibrated, cleaned, and screened event files using the {\it nupipeline} script. The source and background spectra were then extracted from these cleaned event files using the {\it nuproducts} script. To extract both source and background spectra we took circular regions of similar radii (30$^{\prime\prime}$). The source region was centred at the source position and the background region was on the same focal plane module but away from the source contamination. We rebinned the {\it NuSTAR} spectra using \grppha \ routine to have at least 25 counts per spectral bin. 

Additionally, we searched for 3C 273 data in the {\it XMM-Newton} data archive. We found 10 {\it XMM-Newton} observations of 3C 273 simultaneous to {\it NuSTAR} observations, as listed in Table \ref{table:obslog}. To reduce {\it XMM-Newton} data of 3C 273, we used the Science Analysis System (SAS v. 21.0.0) and followed the standard procedures\footnote{\url{https://www.cosmos.esa.int/web/xmm-newton/sas}}. We limited our analysis to the data from the European Photon Imaging Camera (EPIC) pn detector
which is the most sensitive and least impacted by pile-up effects. We started with reprocessing the observation data files (ODFs) to generate the calibrated and concatenated EPIC pn event lists. We then filter out the periods of high background flares. We chose a circular region of 40$^{\prime\prime}$ centred on the source to extract the X-ray spectrum. A background spectrum is also extracted using a circular region of a similar radius from a source-free region. We checked all the observations for pile-up using the task \epatplot \ and corrected the affected observations by removing a region of radius 7.5$^{\prime\prime}$ from the core of the source PSF. Finally, we rebinned the X-ray spectra so as to have at least 25 counts for each background-subtracted spectral bin. 
\begin{figure*}
 \centering
    \includegraphics[width=8cm, height=6cm]{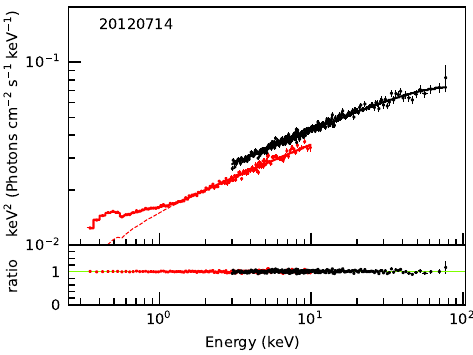} \hfill \includegraphics[width=8cm, height=6cm]{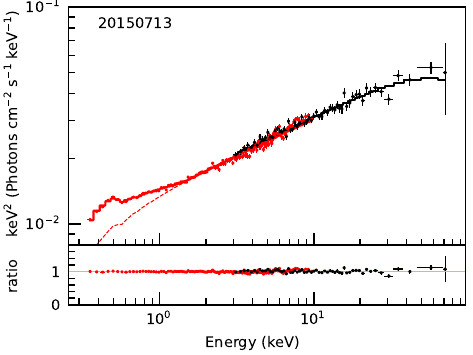}\\
    \caption{\label{fig:SpecJetcaf} Sample plots showing \jetcaf\, model fitted 0.3$-$78.0 keV spectra of 3C\,273 (top panel) and the ratio of data/model (bottom panel). The observation date (yyyymmdd) is mentioned in each plot. Best-fit spectra are rebinned for visual clarity. Plots for the remaining observations are shown in Figure \ref{fig:jetcaf_spec} in the appendix.}
\end{figure*} 

%---------------------------- Results -------------------------------------------------
\section{Results and discussion}\label{sec:result}
We first fitted the 0.3--10 keV {\it XMM-Newton} spectra of 3C 273 using simple power-law (\pow\,) models with Galactic absorption component \tbabs \ using XSPEC version v12.11.0. During the fit, we fixed the value of hydrogen column density to  $1.69\times10^{20}$ cm$^{-2}$ for the Galactic absorption \citep{2016A&A...594A.116H}. The fits were always poor, with $\chi_{r}^2 > 10$  for all the spectra. We illustrate this by plotting the \tbabs(\pow) model fit to the data and the ratio of data/model for each spectrum in Figure \ref{fig:ratio_PL}. The ratio plots show a clear presence of soft excess below 1.5 keV indicating a Seyfert-like feature.

We then fitted joint {\it XMM-Newton} and {\it NuSTAR} X-ray spectra in the full energy range (0.3--78.0 keV)  using the \tbabs(\diskbb+\pow), model. The fitting results are given in Table \ref{table:disk_fit}. The data fitted using this model combination, shown in Figure \ref{fig:2}, returned rather poor fits for all epochs except on MJD 56122. Using the standard relation between \diskbb \ model normalization and the inner edge of the disk ($R_{in}$), one can determine the colour-corrected inner disk radius \citep{Kubota1998},
$R_{in} = [(D/10~\text{kpc})^2 \rm Norm_{diskbb}/\cos i]^{1/2} \kappa^2$ km, where D, $i$, $\rm Norm_{diskbb}$, and $\kappa$ are the source distance, disc inclination, normalization for \diskbb \ component, and the colour correction factor respectively. For the values of D $\sim 750$ Mpc and $i \sim 60^\circ$ \citep{Kriss1999}, M$_{BH} \sim 8\times10^{8}$ M$_{\odot}$ (from this work, see below), $\kappa = 1.6$, and the maximum $Norm_{diskbb}$ value ($\sim 3780$) estimated on MJD 59036, yield $R_{in} \lesssim 1.7\times10^6$ km $\sim 0.07$ $r_S$. Such a low $R_{in}$ indicates that the inner disk extended well within the innermost stable circular orbit or ISCO, which is non-physical. While we can get estimates of accretion disc temperature $T_{in}$ and spectral slope of the coronal emission, those quantities are the end product of underlying fundamental physical quantities i.e. the mass accretion rate \citep{ShakuraSunyaev1973,SunyaevTitarchuk1980,ChakrabartiTitarchuk1995ApJ...455..623C}. Since the origin of soft X-ray excess is unclear yet, we also tested a model with two \gauss\ components combined with a \pow.\ However, the fit is still not satisfactory ($\chi^2_{\rm r}>1.4$). The unsatisfactory results of these model fits motivated us to use an accretion-ejection-based \jetcaf\,model, which was recently developed by \citet{MondalChakrabarti2021}. 

We next performed a broadband (0.3--78.0 keV) X-ray spectral fitting of 3C 273  using simultaneous {\it XMM-Newton} and {\it NuSTAR} observations with the \jetcaf\,model. The \jetcaf\, model fitted parameters are shown in Table \ref{table:JetcafResults}. 
Along with \jetcaf\,model, two \gauss\, components were also included to take into account the soft excess below $\sim1.5$ keV energy range. One component is required between energies $0.2-0.5$ keV of width $0.1-0.4$ keV and another
component at $\lesssim0.1$ keV of width $\sim0.4$ keV. Some representative best fits are shown in \autoref{fig:SpecJetcaf}. The remaining observations are shown in \autoref{fig:jetcaf_spec}. 
When used in XSPEC, the \jetcaf \ model incorporates the total spectrum, including all components,  to fit the observed data. Consequently, the unfolded spectra do not display individual components. However, in theoretical spectra, the individual components can be separated, as demonstrated by \citet{MondalChakrabarti2021}. 

\begin{figure*}
    \centering
    \includegraphics[width=18cm, height=14cm]{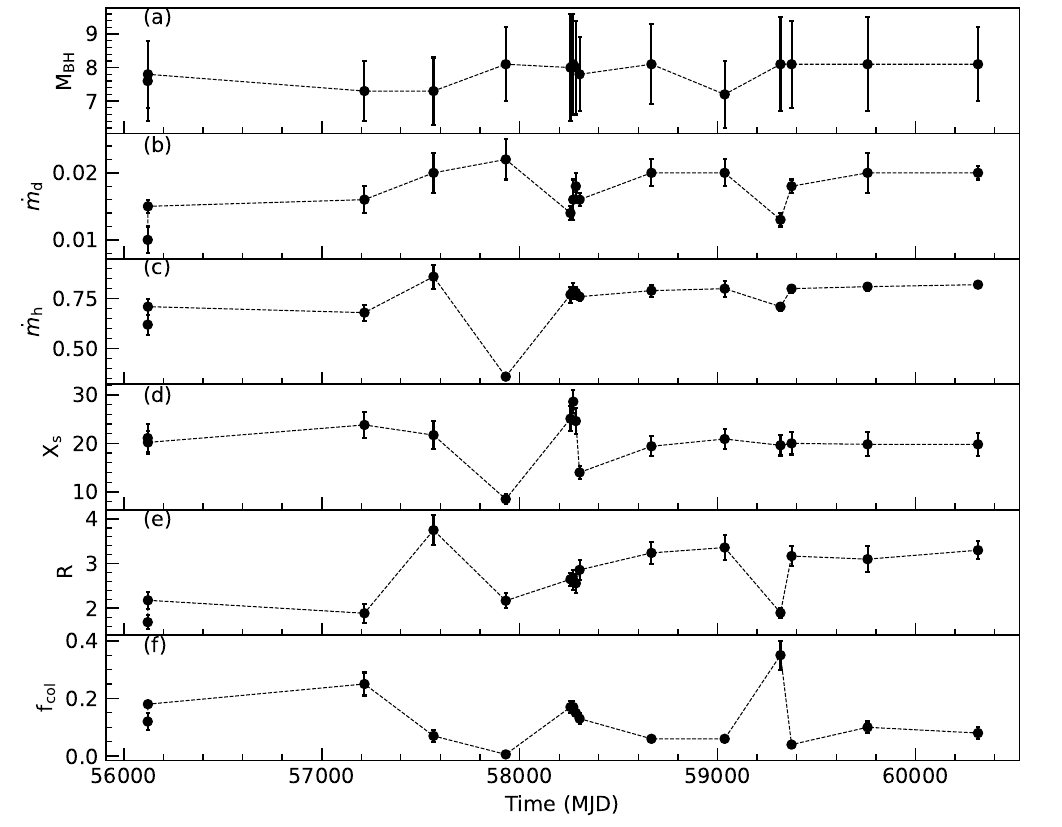}
    \caption{\label{fig:pars_var}Temporal variation of best-fitting \jetcaf\ model parameters: (a) black hole mass, (b) disc accretion rate, (c) halo accretion rate, (d) size of the corona, (e) shock compression ratio, (f) jet collimation factor.}
    
\end{figure*}

The variation of model-fitted parameters with observations is shown in Figure \ref{fig:pars_var}.  During the combined model fitting, we use black hole mass ($M_{BH}$) as a parameter and keep it free from epoch to epoch. The $M_{BH}$ parameter varies between (7.2--8.1)$\times 10^8 M_\odot$ and the error weighted average value comes out to be (7.8$\pm$0.3) $\times 10^8 M_\odot$.  

%%%%%%%%%%%%%%%%%%%%%%%%%%%%xmm+nustar%%%%%%%%%%%%%%%%%%%%%

\begin{table*}
\hspace{-12.0cm}
%\scriptsize
%\small
\centering
\caption{\label{table:JetcafResults} Best fitting  \tbabs*\jetcaf\, model parameters for all observations of 3C\,273. Epochs with simultaneous {\it XMM-Newton} observations showed excesses below $\sim1.5$ keV, which required one Gaussian component between energies 0.2-0.5 keV of width 0.1-0.4 keV and another component at $\lesssim 0.1$ keV of width $\sim 0.4$ keV. The fluxes for the jet (F$_{\rm jet}$) and the corona (F$_{\rm corona}$) are in units of 10$^{-10}$ erg cm$^{-2}$ s$^{-1}$. Here `--' in F$_{jet}$ denotes the lowest normalization epoch, where the total flux is taken to be the coronal flux.}
%\scalebox{1.2}{
%\begin{center}
\resizebox{\textwidth}{!}{\begin{tabular}{cccccccccccc}
\hline
OBS date.& MJD  &$M_{\rm BH}$ &$\dot m_{\rm d}$ & $\dot m_{\rm h}$ & $X_{\rm s}$ & R &$f_{\rm col}$&$\chi_{\rm r}^2$ & F$_{\rm jet}$ & F$_{\rm corona}$ \\
	     & & $(\times 10^8 M_\odot$)&$(\dot m_{\rm Edd}$)&$(\dot m_{\rm Edd}$)&$(r_{\rm S}$)& &  & \\
\hline
2012-07-13 &56121 &$7.6\pm1.2 $&$0.010\pm0.002$&$0.62\pm0.05$ &$21.1\pm2.9 $&$1.69\pm0.16 $&$0.12\pm0.03 $&1.0 	& $2.24\pm0.07$ &$ 0.32\pm0.01$ \\
2012-07-14 &56122 &$7.8\pm1.0 $&$0.015\pm0.001$&$0.71\pm0.04$ &$20.2\pm2.4 $&$2.18\pm0.19 $&$0.18\pm0.01 $&1.0 	& $2.33\pm0.12$ & $0.20\pm0.01$ \\
2015-07-13 &57216 &$7.3\pm0.9 $&$0.016\pm0.002$&$0.68\pm0.04$ &$23.8\pm2.6 $&$1.89\pm0.21 $&$0.25\pm0.04 $&1.2 	& $1.67\pm0.11$ & $0.15\pm0.01$ \\
2016-06-26 &57565 &$7.3\pm1.0 $&$0.020\pm0.003$&$0.86\pm0.06$ &$21.7\pm2.9 $&$3.75\pm0.33 $&$0.07\pm0.02 $&1.3 	& $3.80\pm0.11$ &$ 0.73\pm0.02$ \\
2017-06-26 &57930 &$8.1\pm1.1 $&$0.022\pm0.003$&$0.36\pm0.01$ &$ 8.5\pm1.0 $&$2.17\pm0.17 $&$0.006\pm0.001$&1.1	& $1.74\pm0.05$ & $0.69\pm0.02$ \\
2018-05-19 &58257 &$8.0\pm1.6 $&$0.014\pm0.001$&$0.77\pm0.04$ &$25.1\pm2.6 $&$2.65\pm0.15 $&$0.17\pm0.02 $&1.1 	& $1.68\pm0.04$ & $0.46\pm0.01$ \\
2018-06-02 &58271 &$8.1\pm1.5 $&$0.016\pm0.003$&$0.79\pm0.04$ &$28.6\pm2.4 $&$2.64\pm0.22 $&$0.17\pm0.02 $&1.1 	& $1.45\pm0.03$ & $0.94\pm0.02$ \\
2018-06-15 &58284 &$8.0\pm1.4 $&$0.018\pm0.002$&$0.78\pm0.03$ &$24.6\pm2.7 $&$2.56\pm0.21 $&$0.15\pm0.01 $&1.0 	& $1.30\pm0.03$ & $0.41\pm0.01$ \\
2018-07-04 &58303 &$7.8\pm1.1 $&$0.016\pm0.001$&$0.76\pm0.02$ &$14.0\pm1.3 $&$2.86\pm0.23 $&$0.13\pm0.02 $&1.0 	& $1.50\pm0.06$ & $0.24\pm0.01$ \\
2019-07-02 &58666 &$8.1\pm1.2 $&$0.020\pm0.002$&$0.79\pm0.03$ &$19.4\pm2.1 $&$3.24\pm0.24 $&$0.06\pm0.01 $&1.1 	& $1.24\pm0.04$ & $0.35\pm0.01$ \\
2020-07-06 &59036 &$7.2\pm1.0 $&$0.020\pm0.002$&$0.80\pm0.04$ &$20.9\pm2.0 $&$3.36\pm0.28 $&$0.06\pm0.01 $&1.1 	& $1.62\pm0.04$ &$ 0.40\pm0.01$ \\
2021-04-14 &59318 &$8.1\pm1.4 $&$0.013\pm0.001$&$0.71\pm0.02$ &$19.6\pm2.1 
$&$1.90\pm0.11 $&$0.35\pm0.05 $&1.1 	& $0.17\pm0.00$ &$ 1.39\pm0.01$ \\
2021-06-09 &59374 &$8.1\pm1.3 $&$0.018\pm0.001$&$0.80\pm0.02$ &$20.0\pm2.3 $&$3.17\pm0.22 $&$0.04\pm0.01 $&1.1 	& $1.06\pm0.03$ & $0.44\pm0.01$ \\
2022-06-28 &59758 &$8.1\pm1.4 $&$0.020\pm0.003$&$0.81\pm0.02$ &$19.8\pm2.5 $&$3.10\pm0.29 $&$0.10\pm0.02 $&1.1 	& $1.24\pm0.03$ &$ 0.38\pm0.01$ \\
2023-04-12 &60046 &$7.9\pm0.9 $&$0.017\pm0.003$&$0.76\pm0.03$ &$23.9\pm2.6$ &$4.69\pm0.53 $&$0.03\pm0.01 $&1.1	& -- & $1.45\pm0.01$ \\
2024-01-07 &60316 &$8.1\pm1.1 $&$0.020\pm0.001$&$0.82\pm0.01$ &$19.8\pm2.4 $&$3.30\pm0.20 $&$0.08\pm0.02 $&1.0 	& $1.15\pm0.03$ & $0.49\pm0.01$ \\
\hline
\end{tabular} }
 %       }
%\end{center}
\end{table*}

Different estimates of the black hole mass of 3C 273 span a broad range of values that were obtained using various methods. Using the reverberation mapping (RM) technique, \cite{Laor1998} estimated a BH mass of 7.4 $\times 10^8 M_\odot$. \cite{Kriss1999} applied the accretion disk models to broadband (UV to X-ray) data of 3C 273 to get $M_{BH}$ in a range (7.1--12) $\times 10^8 M_\odot$. 
\cite{kaspi2000} calculated a black hole mass of 2.35$_{-0.33}^{+0.37}$ $\times 10^8 M_\odot$ from the RMS spectra and  5.50$_{-0.79}^{+0.89}$ $\times 10^8 M_\odot$ from average spectra using RM of Balmer lines. \cite{PaltaniTurler2005} obtained a  substantially higher value of M$_{BH}$=6.6$_{-0.9}^{+1.6}$ $\times 10^9 M_\odot$ using RM of broad UV  (Ly$\alpha$ and C $IV$) lines. A black hole mass of only 2.6$\pm$1.1 $\times 10^8 M_\odot$ was measured using the GRAVITY observations \citep{gravity2018}. 
 In a recent study, \cite{2019ApJ...876...49Z} estimated a black hole mass of 4.1$_{-0.4}^{+0.3}$ $\times 10^8 M_\odot$ using a H$\beta$ RM campaign carried out from 2008 to 2018.  Our $M_{BH}$ value lies within this range.

Among the other \jetcaf\ model parameters, the disc mass accretion rate (\md) varies between $0.010\pm0.002$ to $0.022\pm0.003$ $\dot m_{\rm Edd}$, where $\dot m_{\rm Edd}$ is the Eddington accretion rate.  However, the halo, or sub-Keplerian rate (\mh) varies between $0.36\pm0.01$ to $0.86\pm0.06$ $\dot m_{\rm Edd}$, with its value always higher than \md. This implies that the spectra are dominated by the hot flow, i.e., of a hard nature. As both the mass accretion rates varied from epoch to epoch, the size of the dynamic corona ($X_s$) also changes significantly from $8.5\pm1.0$ to $28.6\pm2.4$ $r_S$. 
It is worth noting that during MJD 57930 both $X_s$ and \mh\, were at minima while and \md\ was at its maximum. Such a negative correlation is expected in \jetcaf\, framework as the increased accretion rate increases the amount of cooling of the corona and therefore the shock moves inward \citep{ChakrabartiTitarchuk1995ApJ...455..623C,MondalChak2013MNRAS.431.2716M}.

The shock compression ratio ($R$ = velocity of the flow inside the corona/outside the corona) also changes significantly, from $1.9\pm0.2$ to $3.8\pm0.3$. The jet collimation factor ($f_{\rm col}$) variation ranges from a well-collimated outflow (the lowest value; 0.006) to almost a wind-like outflow (the highest value; 0.35). When  \md\, is high and $f_{col}$ is near a minimum, we can understand why the outflow is weakest: a cooler corona may not have enough pressure to drive the outflow \citep{Chakrabarti1999}. %The $\Gamma$ index indeed showed variation compared to its standalone case, this implies that the harder component of the spectra, which can originate from the extended jet/outflow, is variable. 

Overall, the data fitted quite well with this model, which returned reduced $\chi^2 \simeq 1-1.1$ except for MJDs 57216 and 57565.  The best \jetcaf\,model fits to the $0.3-78.0$ keV spectra of 3C 273 as well as the ratios of data/model are shown in Figure \ref{fig:SpecJetcaf}. The best-fitted spectra are rebinned for visual clarity, which does not affect the fitted parameters.  After achieving the best fit, we have estimated the corona flux (F$_{\rm corona}$) and jet flux (F$_{\rm jet}$) using the lowest model normalization method \citep[see][]{Jana2017,MondalEtal2022A&A...663A.178M}. The best-fitting model gives the total flux (F$_{\rm total}$) and after replacing the best-fitted model norm by the lowest value obtained from MJD 60046 for this source, gives the F$_{\rm corona}$. Subtracting F$_{\rm corona}$ from F$_{\rm total}$ yields F$_{\rm jet}$. Therefore, on MJD 60046 there is essentially no jet contribution. On this observation date, the $R$ parameter was maximal ($\sim5$),  yielding a much lower outflow rate (which is only a function of $R$) and, therefore a much lower flux, which is consistent with the observed spectrum.
All \jetcaf\, model parameters and their variation with MJD are shown in Figure \ref{fig:pars_var}.
It is to be noted that \jetcaf\, model includes the base of the jet, extending from above the corona up to the sonic surface; however, large-scale highly collimated jets \citep{MarshallEtal2001ApJ...549L.167M} can not be taken into account using this model. The radius of the sonic surface is estimated using the relation $r_c=f_0 X_s/2$ \citep{Chakrabarti1999}, where $f_0=R^2/R-1$, which yields $r_c\sim 2.5\times X_s$ (in $r_S$) for the strong shock case.

In $\gamma$-rays, 3C 273 evinces a blazer-like beamed emission component produced by the relativistic jets, so some of the flux is Doppler boosted. Such emission would include photons produced in the jet by the inverse Compton process off of thermal and synchrotron X-ray photons from the disc and corona.
 The present version of the \jetcaf\, model does not include relativistic beaming effects, and the $F_{\rm jet}$ is calculated as the total jet flux which may include contributions from other physical processes. The estimated $F_{\rm jet}$ is significantly higher than  $F_{\rm corona}$  in almost all epochs. Earlier \citet{Grandi2004} reported quite similar results while comparing the jet flux with the Seyfert (non-jetted) type flux using reflection-based models. Moreover, previous estimates of $F_{\rm jet}$ in other jetted sources also showed $F_{\rm jet}$ is higher by a factor of few or comparable to $F_{\rm corona}$  when the jet was active or moderately weak, using the \tcaf\, model \citep{Jana2017,MondalEtal2022A&A...663A.178M}. However, for the present source, the jet flux was always higher, which can be due to the effect of Doppler boosting. In such a case, the estimated flux can be used as a tool to estimate the Doppler boosting factor $\delta$ \citep[as discussed in][]{BritzenEtal2007A&A...476..759B,HovattaEtal2009A&A...494..527H}. Assuming that the lowest jet flux observed on MJD 59318 is the base flux and any excess above that is due to the Doppler boosting  $\delta^4$. Then the minimum and maximum $\delta$ estimated for the MJD 59374 and 57565 are 1.6 and 2.2 respectively. This also interprets the anti-correlation between $f_{\rm col}$ and $F_{\rm jet}$. Our estimated $\delta$ is in agreement with the lowest limit reported by \citet{AbrahamRomero1999A&A...344...61A}.

As a further consistency check of the analysis method and to verify the jet signature in the observed spectra, we have redone the fitting using only the \tcaf\,model. For all epochs where {\it NuSTAR} data is available only, \tcaf\,returned good fits that are similar to \jetcaf\,model-fitted statistics. The joint broadband spectral fitting with \tcaf\,along with \gauss\,components for soft-excess also returns satisfactory fits, although with marginally higher fit statistics compared to the \jetcaf\,, $\Delta \chi_r^2 \gtrsim 0.1$. The improvement in fit statistics is due to the presence of two additional components in \jetcaf: (1) some excess at the shoulder of the blackbody ($\sim 2-5$ keV); and (2) excess above $\sim 50$ keV. Keeping in mind that 3C 273 is a ``jetted source'', we compared the parameters of both models and found that the \tcaf\,model fits require a higher $\dot m_d$ and $R$ for all epochs. Noticeably, the $R$ values obtained from \tcaf\, fits are high, falling in the $> 4$ range. However, in the \tcaf\,scenario, jets/outflows are significant and are launched for intermediate values of R \citep[$\sim 2-3$;][]{Chakrabarti1999} and when the disk accretion rate is low. These are closely consistent with \jetcaf\,model fitted parameters. The $\dot m_h$ and $X_s$ obtained from both models are nearly similar. This comparison shows that the contribution of the jet to the observed spectra is robust in the model fitting. 

%----------------- Conclusion -----------
\section{Conclusions}\label{sec:conclusion} 
The extremely bright nearby quasar 3C 273 has shown the properties of both jetted AGN and Seyfert galaxies, making it a perfect source for examining the disk-jet interactions in AGN. In this work, we fitted the broadband (0.3--78.0 keV) X-ray data of 3C 273 from {\it XMM-Newton} and {\it NuSTAR} using combined accretion-ejection and jet-based models. We summarize the main findings of our investigation below:
\begin{itemize}
    \item A simple PL fit left significant X-ray excess at energies below $\sim$ 1.5 keV, suggesting emission from the accretion disk-corona system.
    \item The accretion disc model that was added to PL did not yield an acceptable fit. 
    \item The accretion-ejection-based \jetcaf\ model provided the best fits along with sensible accretion flow parameters. 
    \item The value of the model parameter M$_{\rm BH}$ remains consistent for all the observations. The weighted-mean value of M$_{\rm BH}$ is (7.77$\pm$0.30) $\times 10^8 M_\odot$, which falls in the range of M$_{\rm BH}$ values estimated using RM and other accretion based models in the literature. 
    \item A broad range of the $f_{col}$ parameter indicates the presence of both collimated and wind-like outflows from the system.
    \item For all observations \mh>\md\, and $X_s$ was relatively high, a combination associated with the hard spectra from the corona region, and consistent with the PL model fitted $\Gamma$, $<1.8$.

    \item If the estimated jet flux is used to estimate the Doppler boosting factor, we obtain values between 1.6--2.2, consistent with the lowest value found in the literature. 
\end{itemize}

\begin{acknowledgements}
We sincerely thank the referee for their insightful comments and suggestions. The project was partially supported by the Polish Funding Agency National Science Centre, project 2017/26/A/ST9/00756 (MAESTRO 9). AP acknowledges funding from the Chinese Academy of Sciences President’s International Fellowship Initiative (PIFI), Grant No. 2024PVC0088. SM acknowledges the Ramanujan Fellowship (RJF/2020/000113) by DST-SERB, Govt. of India for this research.
\end{acknowledgements}

\bibliographystyle{aa} 
\bibliography{master} 

\begin{appendix} %First appendix
%\FloatBarrier

\section{Plots showing PL fitting to 0.3-10 keV spectra of 3C 373}

\begin{figure*}
    \centering
  \includegraphics[width=8cm, height=6cm]{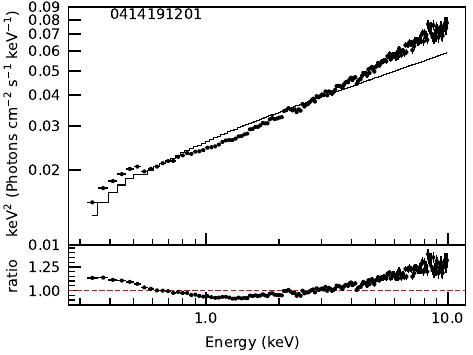} \includegraphics[width=8cm, height=6cm]{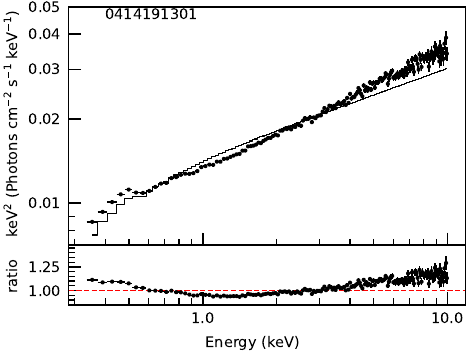}\\
   \includegraphics[width=8cm, height=6cm]{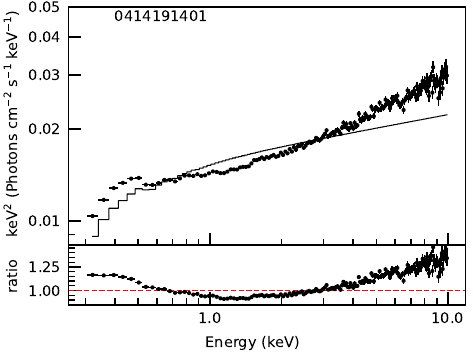} \includegraphics[width=8cm, height=6cm]{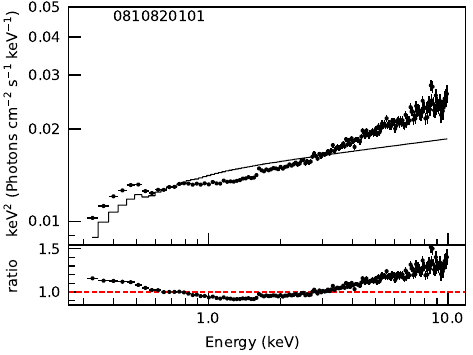}\\
    \includegraphics[width=8cm, height=6cm]{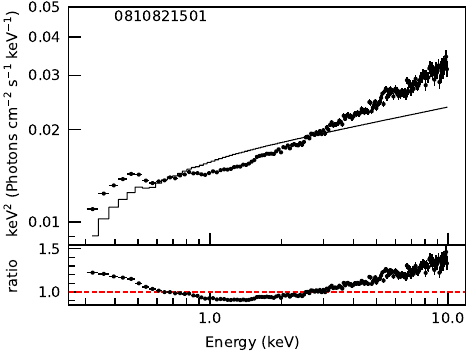} \includegraphics[width=8cm, height=6cm]{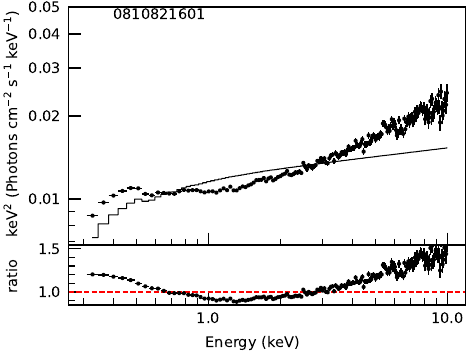}\\
 \includegraphics[width=8cm, height=6cm]{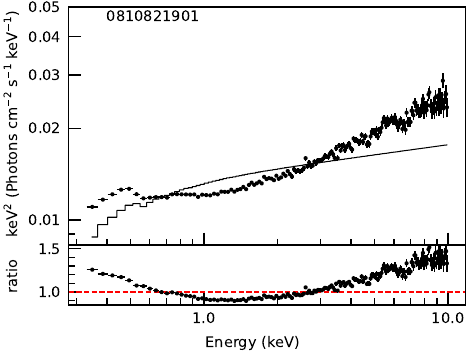} \includegraphics[width=8cm, height=6cm]{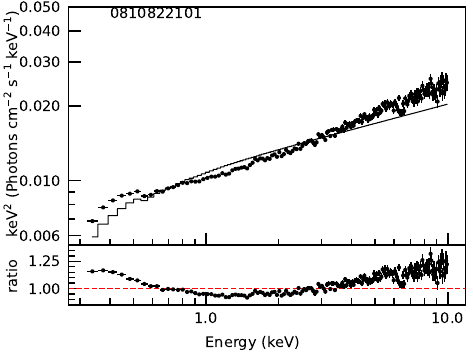}\\
    \caption{\label{fig:pl_spec} A simple PL fitted 0.3-10 keV spectra of 3C 273. The ratio of data/model is plotted in the bottom portion of each panel.
    }
    
\end{figure*}

\section{Plots showing \tbabs(\diskbb+\pow) fitting to 0.3-78 keV spectra of 3C 273}
%\FloatBarrier
\begin{figure*}
     \includegraphics[width=8cm, height=6cm]{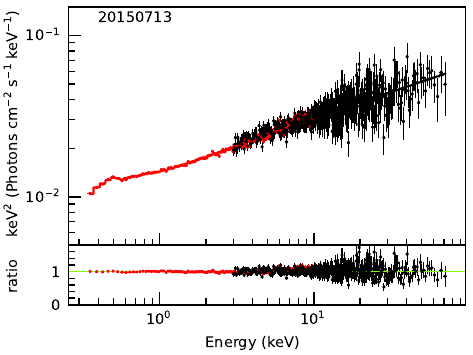}
      \includegraphics[width=8cm, height=6cm]{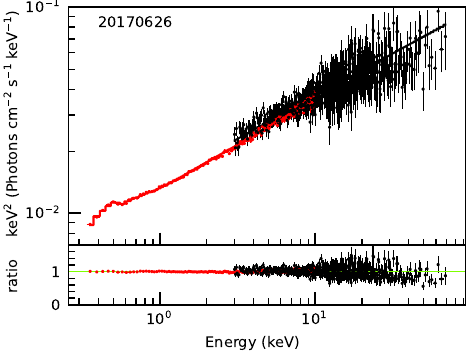}\\
      \includegraphics[width=8cm, height=6cm]{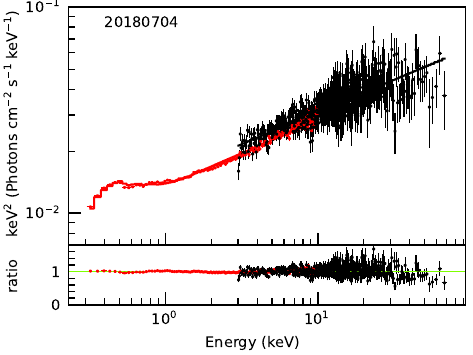}
      \includegraphics[width=8cm, height=6cm]{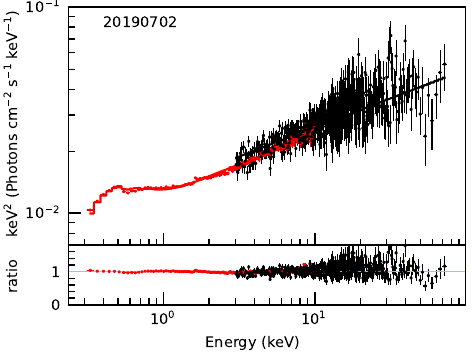} \\
        \includegraphics[width=8cm, height=6cm]{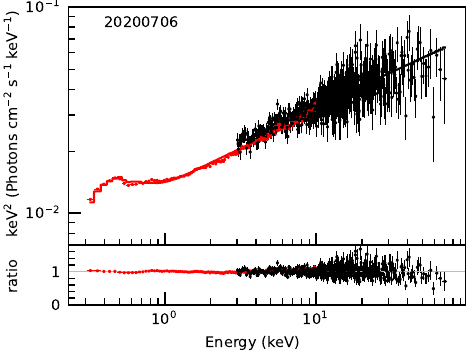}
     \includegraphics[width=8cm, height=6cm]{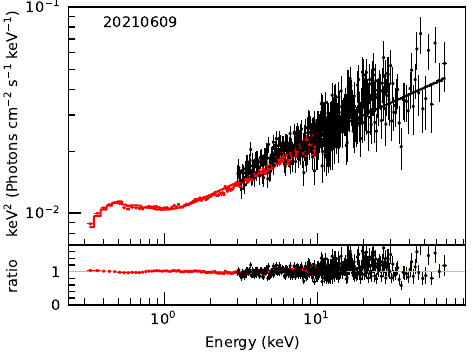} \\
     \includegraphics[width=8cm, height=6cm]{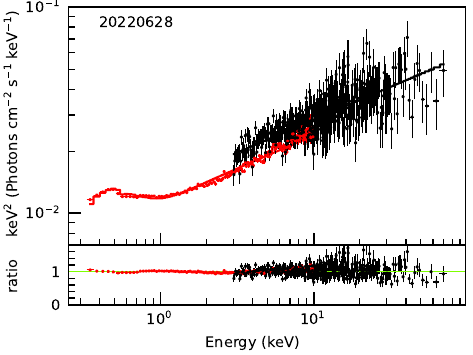}
      \includegraphics[width=8cm, height=6cm]{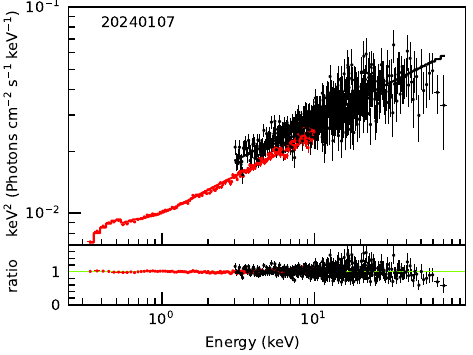}\\
          \caption{\label{fig:diskbb+pl_spec} The \tbabs(\diskbb+\pow) model fitted 0.3-78 keV spectra of 3C 273. The ratio of data/model is plotted in the bottom portion of each panel.
    }
\end{figure*}

\section{Plots showing \jetcaf \ model fitted 0.3-78.0 keV spectra of 3C 273}
%\FloatBarrier
\begin{figure*}
 \centering
    \includegraphics[width=8cm, height=6cm]{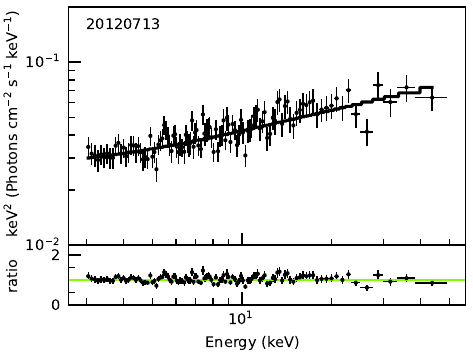}\includegraphics[width=8cm, height=6cm]{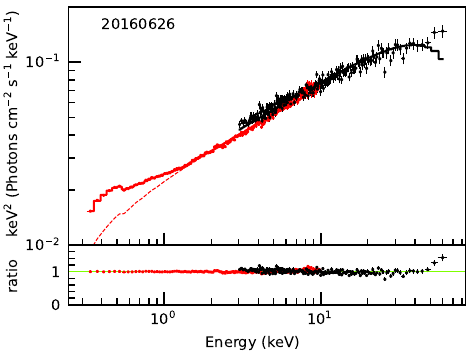} \\
     \includegraphics[width=8cm, height=6cm]{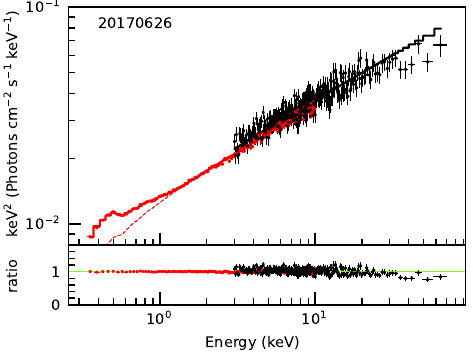}\includegraphics[width=8cm, height=6cm]{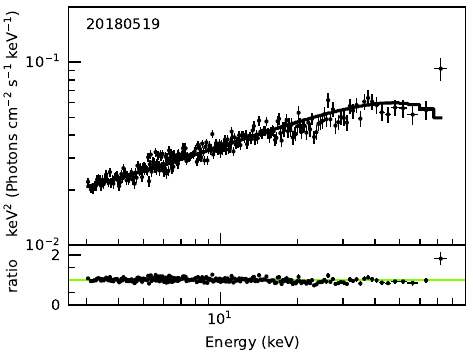} \\
      \includegraphics[width=8cm, height=6cm]{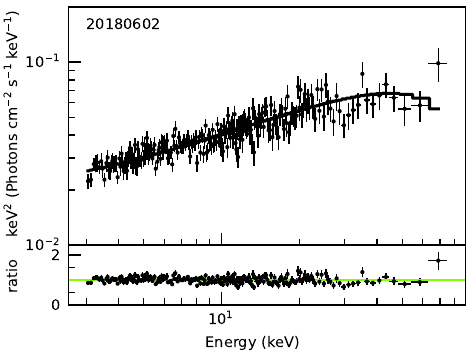}\includegraphics[width=8cm, height=6cm]{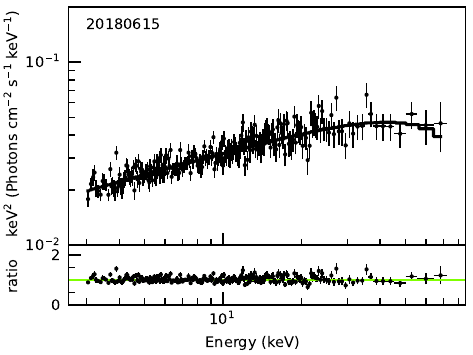} \\
       \includegraphics[width=8cm, height=6cm]{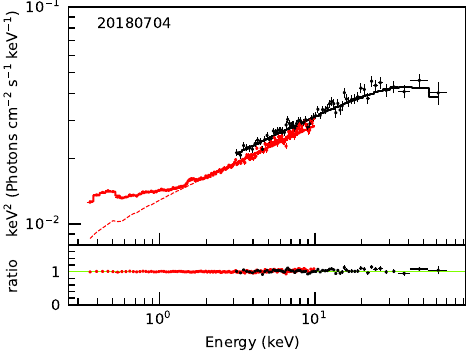}\includegraphics[width=8cm, height=6cm]{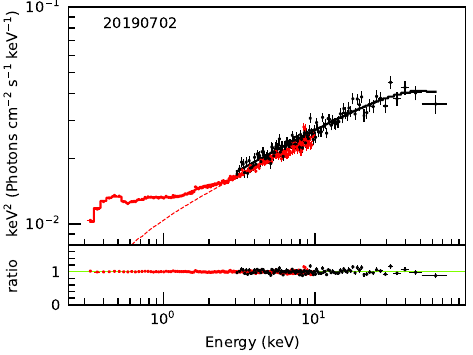} 
\caption {\label{fig:jetcaf_spec} \jetcaf\,model fitted 0.3$-$78.0 keV spectra of 3C\,273. The ratio of data/model is plotted in the bottom portion of each panel.}
\end{figure*} 

\begin{figure*}
\addtocounter{figure}{-1}
 \centering

      \includegraphics[width=8cm, height=6cm]{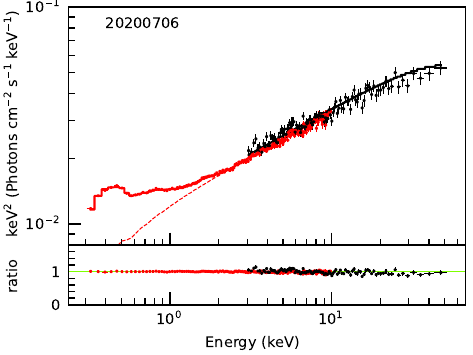}\includegraphics[width=8cm, height=6cm]{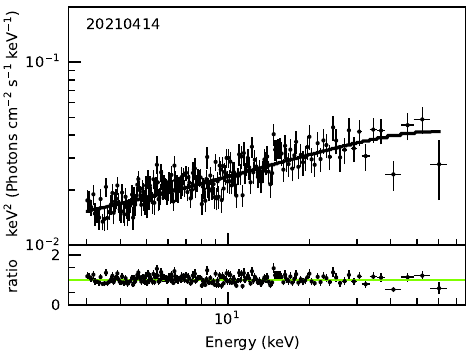}\\
    \includegraphics[width=8cm, height=6cm]{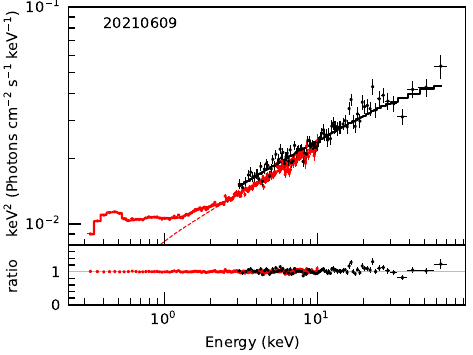}\includegraphics[width=8cm, height=6cm]{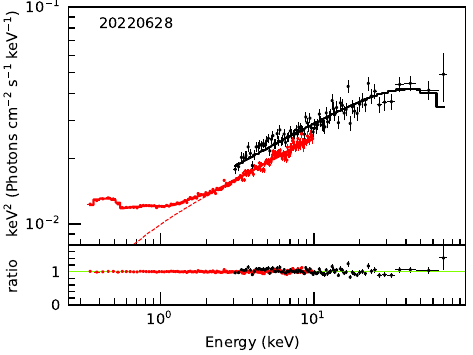} \\
     \includegraphics[width=8cm, height=6cm]{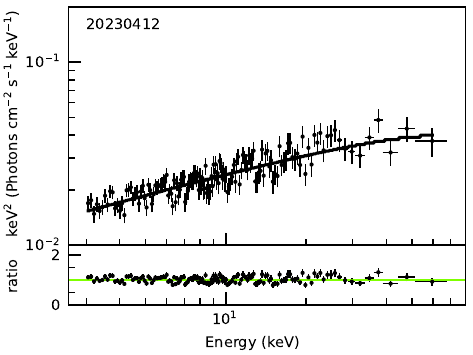}\includegraphics[width=8cm, height=6cm]{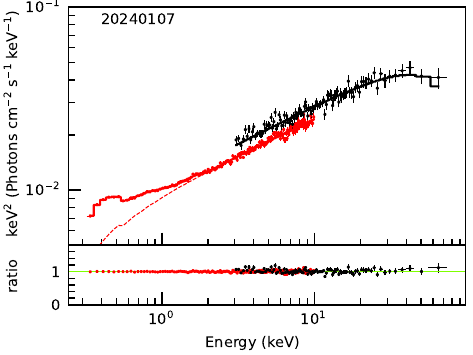} 
\caption{Continued.}

\end{figure*} 

\end{appendix}
\end{document}